\documentclass[conference]{IEEEtran}

\usepackage{amsmath,amssymb,amsfonts}
\usepackage{algorithmic}
\usepackage{graphicx}
\usepackage{textcomp}
\usepackage{xcolor}
\usepackage{algorithmic}
\usepackage[T1]{fontenc}
\usepackage{algorithm}
\DeclareMathOperator*{\argmax}{arg\,max}
\DeclareMathOperator*{\argmin}{arg\,min}
\usepackage{cases}
\usepackage{cite} 
\usepackage{amsthm}

\newtheorem{lemma}{Lemma}
\newtheorem{remark}{Remark}

\usepackage{subfigure}

\begin{document}

\title{Decentralized Network Topology Design for Task Offloading in Mobile Edge Computing}

\author{\IEEEauthorblockN{Ke Ma}
\IEEEauthorblockA{\textit{ Department of Electrical and Computer
Engineering} \\
\textit{University of California, San Diego and San Diego State
University}\\
La Jolla, USA \\
kem006@ucsd.edu}
\and
\IEEEauthorblockN{Junfei Xie}
\IEEEauthorblockA{\textit{ Department of Electrical and Computer
Engineering} \\
\textit{San Diego State University}\\
 San Diego, USA \\
jxie4@sdsu.edu}
}

\maketitle
\begin{abstract}
The rise of delay-sensitive yet computing-intensive Internet of Things (IoT) applications poses challenges due to the limited processing power of IoT devices. Mobile Edge Computing (MEC) offers a promising solution to address these challenges by placing computing servers close to end users. Despite extensive research on MEC, optimizing network topology to improve computational efficiency remains underexplored. Recognizing the critical role of network topology, we introduce a novel decentralized network topology design strategy for task offloading (DNTD-TO) that jointly considers topology design and task allocation. Inspired by communication and sensor networks, DNTD-TO efficiently constructs three-layered network structures for task offloading and generates optimal task allocations for these structures. Comparisons with existing topology design methods demonstrate the promising performance of our approach.   

\end{abstract}


\section{Introduction}
With the advancement in Internet of Things (IoT) technology, many delay-sensitive yet computing-intensive applications have emerged, such as automotive driving, face recognition and virtual reality\cite{yang2018communication}. However, IoT devices typically have limited computing power, making it difficult to meet the demands of these tasks. Centralized cloud computing is traditionally used to process these tasks, but offloading them to the remote cloud can cause significant transmission delays that degrade user Quality of service (QoS). To address this issue, Mobile Edge Computing (MEC) solutions were introduced\cite{hu2015mobile}, where MEC places servers closer to end users at the network edge, such as at base stations, to reduce transmission delays. 

To better serve end users, extensive research has been conducted in the field of MEC, addressing various design aspects such as system deployment, task offloading, resource allocation, mobility management, and privacy and security \cite{yang2023survey}. Nevertheless, little attention has been given to optimizing network topology to enhance computational efficiency. Most existing studies primarily focus on offloading tasks from users to one or more nearby MEC servers within communication range \cite{linderoth2000metacomputing, 9998108}. A few studies  \cite{qi2024bridge, liu2023joint} have explored offloading tasks to servers multiple hops away, but these designs did not consider the impact of network topology. In our prior work \cite{10622383}, we proposed a multi-layered task offloading framework and demonstrated that computational efficiency can be improved by leveraging layered network structures. We also showed that computational efficiency is influenced not only by the computing and communication characteristics of the servers but also by their network topology. In this paper, we aim to further investigate the joint design of layered network structure and task allocation.

Layered structures have been widely used in communication and practical sensor networks due to energy efficiency and network management simplicity \cite{priyadarshi2024energy}. In these networks, a base station is typically present alongside several clusters of sensors or communication devices, with each cluster comprising a Cluster Head (CH) and multiple Cluster Members (CMs). The CH collects data from its CMs and transmits it to the base station. Methods for selecting CHs and CMs can be broadly categorized into two types:  \textit{cluster-based} \cite{ heinzelman2000energy, 1045297, koltsidas2011game, xie2013efficient} and \textit{grid-based} \cite{chen2009unequal, logambigai2018energy, chiang2014cycle, farman2016grid}. In cluster-based methods, CHs are selected directly based on certain criteria. In contrast, grid-based methods first divide the network into grids, and then select CHs within each grid. Despite the widespread use of layered structures in communication and sensor networks, their design for task offloading remains underdeveloped.


In this paper, we introduce a decentralized network topology design strategy for task offloading (DNTD-TO). We explore three-layered network structures, similar to those commonly used in communication and sensor networks, to facilitate computing. In this setup, tasks are offloaded from the root node (referred to as master) to servers in the second layer (CHs), which then distribute the tasks to their child nodes in the third layer (CMs). To select CHs and CMs, our strategy iterates through two nested phases: a \textit{local cluster formation} phase, where each server within master's communication range selects CMs in a decentralized manner, and a \textit{cluster selection} phase where the master selects CHs and their associated CMs. The selection of CHs and CMs is based on servers' task processing capacities and their potential to enhance computational efficiency. 
Optimal task allocation is integrated into every step of the selection process. 


The rest of the paper is organized as follows. Sec. \ref{sec:modeling} covers system modeling
and problem formulation. Our approach and simulation studies are presented in Sec. \ref{sec: method} and Sec. \ref{sec:simulation}, respectively. 
Sec. \ref{sec:conclusion} concludes the paper.

\section{System Modeling and Problem Formulation} \label{sec:modeling}
Consider an MEC system with $N$ servers
 scattered in an open area,   
each communicating wirelessly only with nearby servers within a uniform communication range $\xi$. 
Suppose one of the servers, referred to as the \textit{master} (e.g., a server located at or near a base station), receives computation task processing requests from end users with a total task size of $Y$. To process tasks efficiently, the master allocates tasks, which are assumed to be arbitrarily decomposable, to its neighbors, referred to as CHs. The CHs then decide whether to further offload these tasks to their own neighbors, referred to as CMs. 
Since simply offloading tasks to all neighbors may not yield optimal performance, this study investigates the joint optimization of network topology and task allocation. 

\subsection{System Modeling}
The MEC system is modeled as a graph $G = \{\mathcal{V}, \mathcal{E}\}$, where $\mathcal{V} = \{0,1,\ldots,N-1\}$ denotes the set of servers, with the master server labeled as $0$. $\mathcal{E}$ represents the set of edges. An edge between two servers, $i$ and $j$, is established when they are within each other's communication range, meaning their Euclidean distance, denoted as $d_{ij}$, satisfies $d_{ij} \leq \xi$. The connectivity is described by the adjacency matrix $A$, where each element $a_{ij}$ is equal to $1$ if there is an edge between server $i$ and server $j$, and $0$ otherwise. 

Additionally, define $\mathcal{N}_i = \{j|a_{ij} = 1, \forall j\in\mathcal{V}\}$, where $i\in\mathcal{V}$, as the set of neighbors of server $i$ that are within its communication range. 
By strategically selecting CHs from the set $\mathcal{N}_0$ to receive tasks offloaded from the master, and assigning CMs from the set $\mathcal{N}_i$ to each CH $i$ to handle tasks offloaded from CH $i$, the master, CHs, and CMs form a three-layer tree topology, denoted as $\mathcal{T}$. We assume that each CM can belong to only one CH. 

To describe data transmission between two servers, we adopt the model in 
\cite{wang2019joint}. Specifically, suppose the servers communicate using the Orthogonal Frequency Division Multiplexing protocol \cite{you2016energy}, 
where the bandwidth is equally divided when a server transmits data simultaneously to its connected servers via orthogonal channels. For simplicity of analysis, we assume the channels are free from interference. If a server $i$ (e.g., a CH)
transmits data to its $m$ neighboring servers (e.g., associated CMs) simultaneously, 
the data transmission rate for the link between server $i$ and its $j$-th neighbor is given by
$    R_{ij} = \frac{B}{m}log_2(1+\frac{\beta_{ij}}{d_{ij}^2})$, 
where $B$ (MHz) denotes the total bandwidth, and $\beta_{ij}$ represents the signal-to-noise (SNR) ratio. 

The computation model from \cite{you2016energy} is used in this study to describe the task processing time. Let $b$ represent the number of CPU cycles required to compute 1 bit of data, and $f_i$ (MHz)  denote the computation capacity of server $i$. Given a task of size $y$ (Gbits), the time taken by server $i$ to process the task  is 
   $ T_i^{comp} = y\gamma_i$, 
where $\gamma_i = \frac{b}{f_i}$ (s/Mbits) indicates the time taken by server $i$ to process one Gbit of data.

\subsection{Problem Formulation}
In this subsection, we present the mathematical formulation of the problem.

The master selects a set of CHs from set $\mathcal{N}_0$ for task offloading. To describe the selection of CHs, we introduce binary decision variables $o_i$, where $i\in \mathcal{N}_0$. The variable $o_i$ equals 1 if server $i$ is selected as a CH, and 0 otherwise. 

After receiving tasks from the master, the CHs select their CMs for further task offloading. To describe the selection of CMs, we introduce decision variables $\boldsymbol{x}_i = [x_{i1},x_{i2}, \ldots,x_{i|\mathcal{N}_i|}]$, where $i \in \mathcal{N}_0$ and $|\mathcal{N}_i|$ finds the cardinality of the set $\mathcal{N}_i$. $\boldsymbol{x}_i \in \mathbb{R}^{|\mathcal{N}_i|}$ is a binary vector, where its $j$-th element, $x_{ij}$, equals 1 if server $j$ is selected to join the cluster $i$, and 0 otherwise.  The resulting cluster with CH $i \in \mathcal{N}_0$, denoted as $C_i$, is given by $C_i = \{j | x_{ij} = 1, \forall j \in \mathcal{N}_i \}$. 

Additionally, we introduce continuous decision variables $y_i \in \mathbb{R}_{\geq 0}$ to represent the size of the tasks offloaded to server $i$, where $i\in \mathcal{V}$. In case when $i=0$, $y_0$ indicates the size of the tasks processed at the master.  


The time required for each server $i$ to receive its assigned tasks can then be expressed as:
\begin{equation}
    T_{i}^{tran} =  \begin{cases}
    o_iy_i\frac{1}{R_{0i}}, & \text{if } i \in \mathcal{N}_0 \\
    o_jx_{ji}y_i(\frac{1}{R_{ji}} + \frac{1}{R_{0j}}), & \text{if } i \in \mathcal{N}_j, j \in \mathcal{N}_0 \\
     0, & \text{else} \\\end{cases}\label{eq:transmission_time}
\end{equation}
Notably, if server $i$ is not selected for task offloading, the associated transmission delay is 0. 

The total time required for each server $i$ to receive and process the assigned computation tasks is given by  
    $\mathcal{J}_i = T_i^{tran} + T_i^{comp}, i \in \mathcal{V}$, 
where $T_i^{comp} = y_i \gamma_i$. Here, we assume the generated results are small in size, making the transmission delay for sending them back to the master negligible, as is often assumed in existing studies  (e.g.,  \cite{8434285}). The total task completion time can then be written as 
    $\mathcal{J} = \max \mathcal{J}_i,\ \forall i \in \mathcal{V}$. 

The objective of this study is to minimize the total task completion time by jointly optimizing the selection of CHs and CMs, and the task allocation, which is formulated as follows: 
\begin{align*} 
\mathcal{P}_{\text{main}}: & \min_{\{y_i\}_{i\in \mathcal{V}}, \{\boldsymbol{x}_{i}, o_i\}_{i\in \mathcal{N}_0}} \quad  \mathcal{J}\\
    s.t. \quad  & o_i \in \{0,1\}, \forall i \in \mathcal{N}_0 & C1\\
    & x_{ij} \in \{0,1\}, \forall i \in \mathcal{N}_0, j\in \mathcal{N}_i & C2\\
    & 0 \leq y_i \leq Y, \forall i \in \mathcal{V} & C3\\
    & C_i \cap C_j = \emptyset, \forall i, j \in \mathcal{N}_0, i \neq j & C4 \\
    & y_i = \sum_{j \in \mathcal{N}_i}x_{ij}y_j, i\in \mathcal{N}_0 & C5 \\
    & y_0 + \sum_{i \in \mathcal{N}_0} o_i y_i = Y & C6
\end{align*}


Constraints $C1$-$C3$ ensure that the decision variables take valid values. Constraints $C4$
ensure that clusters do not overlap. Constraints $C5$-$C6$ maintain the integrity of task sizes. It should be noted that solving this problem directly is challenging due to the nonlinear constraints $C4$. 

\section{Decentralized Network Topology Design for Task Offloading (DNTD-TO)} \label{sec: method}
In this section, we introduce our approach, DNTD-TO, for solving problem $\mathcal{P}_{\text{main}}$. 
This method consists of two nested phases: 
(1) \textit{Local Cluster Formation (LCF)} phase, 
and (2) \textit{Cluster Selection} phase. 


\subsection{Local Cluster Formation (LCF)}
In this phase, each server  $i \in \mathcal{N}_0$ within the master's communication range selects its CMs in a decentralized manner, forming a candidate cluster with itself as the CH. These candidate clusters will then be examined by the master in the cluster selection phase as detailed in the next subsection. 
Notably, this phase allows a server to join multiple clusters. 


To select CMs, each server $i\in \mathcal{N}_0$ employs a forward selection mechanism, picking CMs from its neighbors $\forall j\in \mathcal{N}_i$ one by one until either no further performance gains can be achieved or all neighbors have been evaluated. In each iteration, the CH adds the neighboring server 
that maximally reduces the task processing time, assuming tasks of any size $y$ are assigned to server $i$. 
This selection is achieved by identifying the neighboring server with the highest processing capacity, defined as the time to receive and process a unit-sized task, and determining if its addition would reduce the overall task processing time. This is challenging, as a neighboring server's processing capacity depends on the number of servers in the cluster due to shared bandwidth. Additionally, determining time savings from adding a server requires solving an optimization problem for task allocation. 
In the following, we first derive the processing capacity, denoted as $\alpha_i$. 


Initially, the cluster $C_i$, with server $i\in\mathcal{N}_0$ as the CH, is empty. Therefore, when a neighboring server is added to the cluster, this server can utilize the entire bandwidth resource. Then, for each server $j\in \mathcal{F} = \mathcal{N}_i$ within $i$'s communication range, its processing capacity $\alpha_j$ can be derived as $\alpha_j=\gamma_j + \frac{1}{Blog_2(1+\beta_{ij}d_{ij}^{-2})}, j \in \mathcal{F}$, where $\mathcal{F}$ denotes the set of neighboring servers that have not been added to the cluster. Nevertheless, in subsequent iterations, as the cluster $C_i$ becomes nonempty, a neighboring server $j\in\mathcal{F}$ can no longer utilize the entire bandwidth resource, and its processing capacity $\alpha_j$ is given by: 

\begin{align}
    \alpha_j = \gamma_j + \frac{|C_i| + 1}{Blog_2(1+\beta_{ij}d_{ij}^{-2})}, j \in \mathcal{F} \label{eq: update ability}
\end{align}
Notably, the processing capacity of the CH $i$ is always $\alpha_i = \gamma_i$ due to local computing. 

To determine whether adding a neighboring server $j$ would reduce the task processing time, we define a 
performance indicator $\mathtt{I}$ that 
compares the minimum task processing time before and after the server is added. 
Specifically, 
consider cluster $C_i$ at the $k$-th iteration, for task $y$ assigned to server $i\in \mathcal{N}_0$, 
the minimum task processing time and the optimal task allocation can be derived by solving the following optimization problem:
\begin{align*}
\mathcal{P}_1:   &\min_{y_l, \forall l\in C_i \cup \{i\}} J \\
    s.t. \quad &\sum_{l\in C_i \cup \{i\}} y_l = y
\end{align*}
where $J = \max_{l\in C_i \cup \{i\}} J_l$ is the task processing time and $J_l = y_l\alpha_l$ is the time required for server $l$ to receive and process its assigned task $y_l$. The performance indicator is then defined as $\mathtt{I} = \frac{J^*_{C_i}}{J^*_{C_i \cup \{j\}}}$, where $J^*_{C_i}$ and $J^*_{C_i \cup \{j\}}$ represent the minimum task processing time obtained before and after adding server $j\in \mathcal{F}$ to cluster $C_i$, respectively. Therefore, if $\mathtt{I} > 1$, adding the server to the cluster will improve performance; otherwise, it will not.

To derive the formula for $\mathtt{I}$, we solve the above optimization problem, which leads to the following lemma, with the proof provided in the Appendix.

\begin{lemma} \label{Thm:theorem 1}
Consider problem $\mathcal{P}_1$. The optimal task allocation, denoted as $y^*_l$, $\forall l\in C_i \cup \{i\}$, satisfy 
$\alpha_iy^*_i = \alpha_ly^*_l$, $\forall l\neq i, l\in C_i$, when $C_i \neq \emptyset$. Consequently, the minimum task processing time is given by $J^* = \alpha_iy^*_i = \alpha_ly^*_l$, $l\in C_i$. In the special case where $C_i = \emptyset$, we have $J^* = \alpha_i y$.  
\end{lemma}

From the above lemma, we can derive that $J^*_{C_i} = y_l^{*}\alpha_l$  and $J^*_{C_i\cup\{j\}} = \bar{y}_l^*\bar{\alpha}_l = \bar{y}_j^*\bar{\alpha}_j$ for $l\in C_i\cup \{i\}$. Here, $\alpha_l$ and $\bar{\alpha}_l$ represent the processing capacities for server $l$ before and after adding server $j$ to cluster $C_i$, respectively, while  $y^*_l$ and $\bar{y}^*_l$ denote the corresponding optimal task allocations. Therefore, we have $\mathtt{I} = \frac{y_l^*\alpha_l}{\bar{y}_l^*\bar{\alpha}_l}$. In this equation, the processing capacities $\alpha_l$ and $\bar{\alpha}_l$ can be readily computed using \eqref{eq: update ability}. Specifically, if $l\in C_i$, $\alpha_l = \gamma_l + \frac{|C_i|}{Blog_2(1+\beta_{il}d_{il}^{-2})}$ and $\bar{\alpha}_l =\gamma_l + \frac{|C_i| + 1}{Blog_2(1+\beta_{il}d_{il}^{-2})}$; otherwise, if $l = i$, $\alpha_i = \bar{\alpha}_i = \gamma_i$.   However, determining the optimal task allocations $y_l^*$ and $\bar{y}_l^*$ by solving $\mathcal{P}_1$ at each iteration is time-consuming. To address this issue, we introduce an iterative method to efficiently compute the values of $\mathtt{I}$, $y_l^*$ and $\bar{y}_l^*$. 

Initially, $C_i$ is empty. Hence, we can easily obtain that $y_i^{*} = y$ before adding server $j$, and $\bar{y}^{*}_i = \frac{\bar{\alpha}_jy}{\bar{\alpha}_i + \bar{\alpha}_j}$, $\bar{y}^{*}_j = \frac{\bar{\alpha}_iy}{\bar{\alpha}_i+\bar{\alpha}_j}$ after adding server $j$ based on Lemma \ref{Thm:theorem 1}. Thus, $\mathtt{I}^{} = \frac{\bar{\alpha}_i+\bar{\alpha}_j}{\bar{\alpha}_i}$. In subsequent iterations, to derive $\mathtt{I}$, we note the existence of following relationships:
\begin{subnumcases} {\label{eq: yj}}
\mathtt{I} = \frac{y_l^*\alpha_l}{\bar{y}_l^*\bar{\alpha}_l}, l \in C_i \cup \{i\} \label{eq: yj first}\\
\bar{y}_j^* + \sum_{l \in C_i \cup \{i\}} \bar{y}_l^* = y \label{eq: yj second}\\
\sum_{\forall l \in C_i \cup \{i\}} y_l^* = y \label{eq: yj third}
\end{subnumcases} 

which yields
\begin{align}
\bar{y}_j^* & = \sum_{l \in C_i \cup \{i\}}(1 - \frac{\alpha_l}{\mathtt{I}\bar{\alpha_l}})y_l^* \label{eq: yi bar}
\end{align}

Since $\bar{y}_j^*\bar{\alpha}_j =\bar{y}_l^*\bar{\alpha}_l$, $\mathtt{I} = \frac{y_l^*\alpha_l}{\bar{y}_l^*\bar{\alpha}_l}$, and $y_l^*\alpha_l = y_i^*\alpha_i$, we have
\begin{equation}
\bar{y}_j^*  =  \frac{y_l^*\alpha_l}{\bar{\alpha}_j \mathtt{I}} = \frac{y_i^*\alpha_i}{\bar{\alpha}_j \mathtt{I}} \label{eq: relation}
\end{equation}

Combining \eqref{eq: yj third}, \eqref{eq: yi bar}, and \eqref{eq: relation}, we can derive that
\begin{align}
    \mathtt{I} = \begin{cases}\frac{\bar{\alpha}_i+\bar{\alpha}_j}{\bar{\alpha}_i}, & \text{if } k = 0 \\    \frac{\bar{\alpha}_{j}\sum_{l \in C_i \cup \{i\}}\frac{y_l^* \alpha_l}{\bar{\alpha}_l} + \bar{\alpha}_{j}y_i^*}{\bar{\alpha}_j y}, & \text{if } k \geq 1 \end{cases}\label{eq: k}
\end{align}

Since $y^*_l$ was obtained in the previous iteration, i.e., $y^{*(k)}_l = \bar{y}^{*(k-1)}_l$, where superscript $(k)$ indicates the iteration index,  \eqref{eq: k} can be readily computed. Moreover, once $\mathtt{I}$ is obtained, we can derive $\bar{y}_l^*$ by 
$\bar{y}_l^* = \frac{y_l^*\alpha_l}{\bar{\alpha_l}\mathtt{I}}, \forall l \in C_i \cup \{i\}$,  
and $\bar{y}_j^*$ can be obtained by \eqref{eq: yi bar}. Algorithm \ref{Algo: Cluster Member Selection} summarizes the procedure. 
\begin{algorithm}
    \caption{LCF($i, \mathcal{N}_i$, $y$)} \label{Algo: Cluster Member Selection}
    \label{Algo: Cluster Member Selection}
    \begin{algorithmic}[1]
        \STATE $C_i \leftarrow \emptyset$, $\mathcal{F} \leftarrow \mathcal{N}_i$, $\alpha_i \leftarrow \gamma_i$, $\bar{\alpha}_i \leftarrow \gamma_i$, $y^*_i \leftarrow y$;
        \FOR{$k=0\ \text{to}\ |\mathcal{N}_i|-1$}
            \STATE Compute  $\bar{\alpha}_{l}$ using (\ref{eq: update ability}), $\forall l \in \mathcal{F}$;
            \STATE $j \leftarrow \argmin_{l\in \mathcal{F}} \bar{\alpha}_l$; \label{line: alpha select}
            \STATE Compute $\mathtt{I}$ by \eqref{eq: k};
            \IF {$\mathtt{I} > 1$}
                \STATE Compute $\bar{y}_{j}^*$ by (\ref{eq: yi bar});
                \STATE ${y}_l^* \leftarrow \frac{y_l^*\alpha_l}{\bar{\alpha}_l\mathtt{I}}$, $\forall l\in C_i \cup \{i\}$; $y_j^* \leftarrow \bar{y}_{j}^*$; 
                \STATE $C_i \leftarrow C_i \cup \{j\}$; $\mathcal{F} \leftarrow \mathcal{F}\setminus \{j\}$;
                \STATE $\alpha_l \leftarrow \bar{\alpha}_l$,  $\forall l \in C_i$;
            \ELSE
                \STATE Break;
            \ENDIF
        \ENDFOR        \RETURN $C_i$, $\{y^*_l\}_{l\in \{C_i\} \cup \{i\}}$
    \end{algorithmic}
\end{algorithm}

\begin{remark}
Inspired by Algorithm \ref{Algo: Cluster Member Selection}, we can solve the optimization problem $\mathcal{P}_1$ efficiently using an iterative procedure as outlined in Algorithm \ref{Algo: Optimal task partition}, which generates the optimal solution.   
\end{remark}


\begin{algorithm} [h]
    \caption{OptiSolver-$\mathcal{P}_1$($i, C_i, y$)} 
    \label{Algo: Optimal task partition}
    \begin{algorithmic}[1]
        \STATE $\boldsymbol{y}^* \leftarrow \emptyset$, $k \leftarrow 0$;
        \FOR {$\forall j \in C_i$}
            \STATE Compute $\bar{\alpha}_{j}$, $\mathtt{I}$ using (\ref{eq: update ability}) and \eqref{eq: k}, respectively;
            \STATE Calculate $\bar{y}_j^*$ by (\ref{eq: yi bar}); 
            \STATE ${y}_l^* \leftarrow \frac{y_l^*\alpha_l}{\bar{\alpha}_l\mathtt{I}}$, $\forall l\in C_i \cup \{i\}$; $y_j^* \leftarrow \bar{y}_{j}^*$; $\boldsymbol{y}^* \leftarrow \boldsymbol{y}^* \cup \{y_j^*\}$; 
            \STATE $C_i \leftarrow C_i \cup \{j\}$; $\alpha_l \leftarrow \bar{\alpha}_l$,  $\forall l \in C_i$; $k \leftarrow k + 1$;
        \ENDFOR
        \RETURN $\boldsymbol{y}^*$
    \end{algorithmic}
\end{algorithm}

\subsection{Cluster Selection}
In this phase, the master examines the candidate clusters formed in the LCF phase, selects the CHs and their associated CMs, and resolves any overlaps between clusters.

To select CHs, the master evaluates each server $i\in \mathcal{N}_0$ within its communication range, following a procedure similar to that of LCF. Particularly, in each iteration, the master identifies the neighboring server $i\in \mathcal{N}_0$ with the highest processing capacity and adds it to the set of CHs, denoted as $C_0$, if doing so would reduce task processing time. The iteration stops when no further reduction in processing is achievable or when all neighboring servers of the master have been examined.

Unlike the processing capacity defined for individual servers in the LCF phase, the processing capacity of each candidate CH $i \in \mathcal{N}_0$ here is defined as the time required for it and its CMs, as a team, to receive and process a unit-sized task. Specifically, for server $i$, the time to receive a unit-sized task from the master is $\frac{1}{R_{0i}}$. Moreover, once server $i$ receives this task, according to Lemma \ref{Thm:theorem 1}, the minimum time required for it and its CMs to process this task is $\frac{y_i^*\gamma_i}{y}$, where $y_i^*$ is the output of Algorithm \ref{Algo: Optimal task partition}. Therefore, the processing capacity of server $i\in \mathcal{N}_0$ in the GCF phase, denoted as $\eta_i$, is given by $\eta_i = \frac{y_i^*\gamma_i}{y} + \frac{1}{R_{0i}}$. Notably, the processing capacity of the master is $\eta_0 = \gamma_0$. 

Moreover, to assess whether adding server $i$ to the set of CHs, $C_0$, would improve performance, we use the same performance indicator $\mathtt{I}$, which can be computed similarly as detailed in the previous subsection. Particularly, at the $k$-th iteration, we have:
\begin{align}
    \mathtt{I} = \begin{cases}\frac{\bar{\eta}_0+\bar{\eta}_i}{\bar{\eta}_0}, & \text{if } k =0 \\
    \frac{\bar{\eta}_{i}\sum_{l \in C_0 \cup \{0\}}\frac{y_l^* \eta_l}{\bar{\eta}_l} + \bar{\eta}_{i}y_0^*}{\bar{\eta}_i Y}, & \text{if } k \geq 1\end{cases}\label{eq: k2}
\end{align}
where $\eta_l$ and $\bar{\eta}_l$ denote the processing capacities of server $l \in C_0 \cup \{0\}$ before and after adding server $i$ to the cluster $C_0$, and we have
\begin{equation}
\bar{\eta}_l = 
\frac{\bar{y}_l^*\gamma_l}{y} + \frac{|C_0|+1}{Blog_2(1+\beta_{0l}d_{0l}^{-2})}\label{eq:beta}\end{equation}
when $l\in C_0 \cup \{0\}$.
$y^*_l$ and $\bar{y}^*_l$ denote the optimal task allocations before and after adding server $i$. Similarly, the updating equation for $\bar{y}_l^*$ is
 $    \bar{y}_l^* = 
 \frac{y_l^*\eta_l}{\bar{\eta_l}\mathtt{I}}, \forall l \in C_0 \cup \{0\}$,  
and we have:
\begin{equation}
 \bar{y}_i^*  = \sum_{l \in C_0 \cup \{0\}}(1 - \frac{\eta_l}{\mathtt{I}\bar{\eta_l}})y_l^* \label{eq:update2_yi}
\end{equation}
To resolve any overlaps between clusters, the selected CH $i$ and its CMs are ``removed'' from the network at the end of each iteration and do not participate in subsequent iterations. At the start of each new iteration, the unselected neighbors of the master undergo the LCF phase to update their CMs, ensuring that clusters remain distinct and non-overlapping. Algorithm \ref{Algo: Clustering} outlines the core procedure of our approach, which constructs the three-layer tree topology $\mathcal{T}$ and determines the associated optimal task allocation $\{y^*_l\}_{\mathcal{T}}$.

\begin{remark}
The task allocation $\{y^*_l\}_{\mathcal{T}}$ generated by our approach in Algorithm \ref{Algo: Clustering} is optimal for the topology $\mathcal{T}$.
\end{remark}

\begin{algorithm}
    \caption{DNTD-TO($Y$)} 
    \label{Algo: Clustering}
    \begin{algorithmic}[1]
        \STATE $\mathcal{T} \leftarrow \{0\}$, $y\leftarrow Y$, $\mathcal{F} \leftarrow \mathcal{N}_0$, $\beta_0 \leftarrow \gamma_0$, $\bar{\beta}_0 \leftarrow \gamma_0$, $y_0^* \leftarrow Y$;
    \FOR{$k=0\ \text{to}\ |\mathcal{N}_0|-1$}
            \STATE $\{C_l, \{y^*_h\}_{h\in \{C_l\} \cup \{l\}}\} \leftarrow$ LCF($l, \mathcal{N}_l,y$), $\forall l \in \mathcal{F}$; \label{line: local cluster}
            \STATE Compute $\bar{\beta}_l$ using \eqref{eq:beta}, $\forall l \in \mathcal{F}$;
            \STATE $i = \argmin_{l \in \mathcal{F}} \bar{\beta}_l$; 
            \STATE Compute $\mathtt{I}$ using (\ref{eq: k2});
            \IF{$\mathtt{I} > 1$} \label{selection: global}
                \STATE Compute $\bar{y}_{i}^*$ by (\ref{eq:update2_yi});
                \STATE $y_l^* \leftarrow 
 \frac{y_l^*\beta_l}{\bar{\beta_l}\mathtt{I}}, \forall l \in C_0 \cup \{0\}$; $y_i^* \leftarrow \bar{y}_i^*$;
                \STATE $\mathcal{T} \leftarrow \mathcal{T} \cup \{j\}\cup C_j $; $C_0 \leftarrow C_0 \cup \{j\}$; $\mathcal{F} \leftarrow \mathcal{F} \setminus \{j\}$;
                \STATE $\beta_l \leftarrow \bar{\beta}_l$,  $\forall l \in C_0$;
                \STATE $\mathcal{N}_i \leftarrow \{l | a_{il}=1, l \in \mathcal{V} \setminus \mathcal{T} \}$, $\forall i \in \mathcal{F}$;
            \ELSE 
                \STATE Break;
            \ENDIF
        \ENDFOR
        \RETURN $\mathcal{T}$, $\{y^*_l\}_{l\in \mathcal{T}}$
    \end{algorithmic}
\end{algorithm}


\section{Simulation studies} \label{sec:simulation}
In this section, we conduct simulation studies to evaluate the performance of our approach.

\subsection{Experiment Setup}
We randomly generate and distribute servers within a $100\times100$m area, with the exception of the master, which is fixed at the location $(20, 20)$m. 
The computing power of each server $f_i$ is sampled from a uniform distribution over $[0.1, 10]$ MHz, with $b$ set to $1$. The SNR ratio $\beta_{ij}$ is sampled from a uniform distribution over $[30, 40]$dbm. The task size $Y$ is set to $100$Gbits, and the bandwidth $B$ is set to 50MHz. 
To evaluate the performance of our method, we compare it with the following three benchmarks:
\addtolength{\topmargin}{0.06in}
\begin{itemize}
    \item \textbf{Unequal Cluster (Unequal)}\cite{chen2009unequal}: 
    This method selects CHs by comparing servers' computing power;
    if two CHs are within each other's communication range, the server with higher computing power is selected as the CH. CMs join the nearest CH.
    \item \textbf{Leach-C} \cite{heinzelman2002application}: A centralized approach that calculates each server's probability of being a CH based on computing power (originally, energy level was used), with the top servers selected as CHs. CMs join the CH with the highest received signal strength (originally, minimal communication energy was used).
    \item \textbf{LBAS} \cite{osamy2024lbas}: CHs are selected using an iterative approach that considers distance, number of neighbors, and computing power, with higher-scoring servers selected as CHs. A similar procedure is applied to select CMs.
    \item \textbf{Dijkstra's} \cite{10622383}: A tree topology is constructed by identifying the shortest routes with the least transmission delay from the \textit{master} to each server using the Dijkstra's algorithm. Servers more than two hops away are pruned. 
    
\end{itemize}
Notably, these benchmarks only generate the tree topology $\mathcal{T}$. The associated optimal task allocation is derived using the same approach as ours.

\subsection{Experiment Results}

\begin{figure}
  \centering
  \subfigure[]{
    \centering\includegraphics[width=0.45\linewidth]{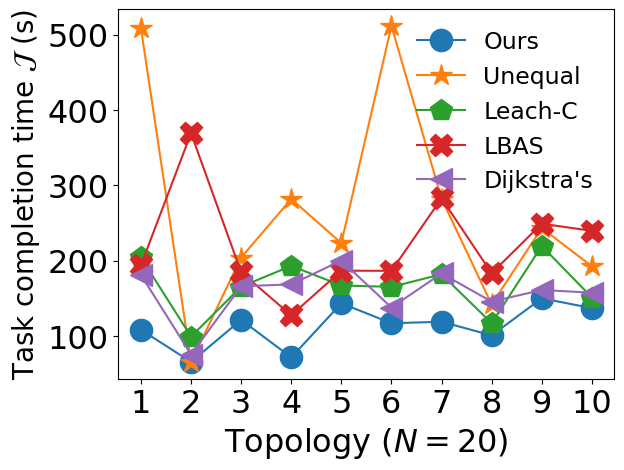}
    \label{fig:Topology_20}}
  \subfigure[]{
    \centering\includegraphics[width=0.45\linewidth]{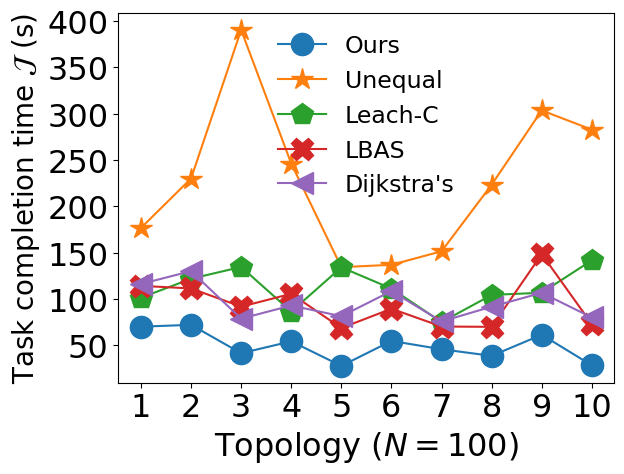}
    \label{fig:Topology_100}}

\caption{Performance comparison across different topologies when (a) $N=20$ and (b) $N=100$.} \label{fig: Topology}
\vspace{-0.3cm}
\end{figure}

In the first experiment, we compare the performance of different approaches across networks of varying sizes and topologies. Specifically, we examine a small-scale network with $N=20$ servers and a large-scale network with $N=100$ servers. For each network size, we randomly generate 10 different topologies by varying server locations, with each server's communication range set to $50$ m. 
As shown in Fig. (\ref{fig: Topology}), our approach outperforms all benchmarks in both small-scale (Fig. \ref{fig:Topology_20}) and large-scale networks (Fig. \ref{fig:Topology_100}). The performance advantage is more pronounced in large-scale networks, due to the increase of servers that can potentially slow down processing if included without careful selection. 
Comparing the two figures, we can see that the task completion time generally decreases with the increase of the number of servers, due to more servers participating in computations.

\begin{figure}
  \centering
  \subfigure[]{
    \centering\includegraphics[width=0.35\linewidth]{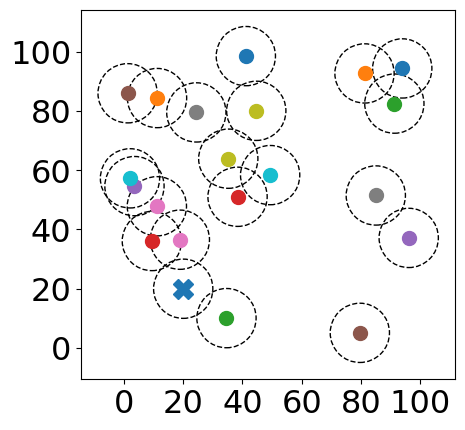}
    \label{fig:COMMRANGE_fig}}
  \subfigure[]{
    \centering\includegraphics[width=0.45\linewidth]{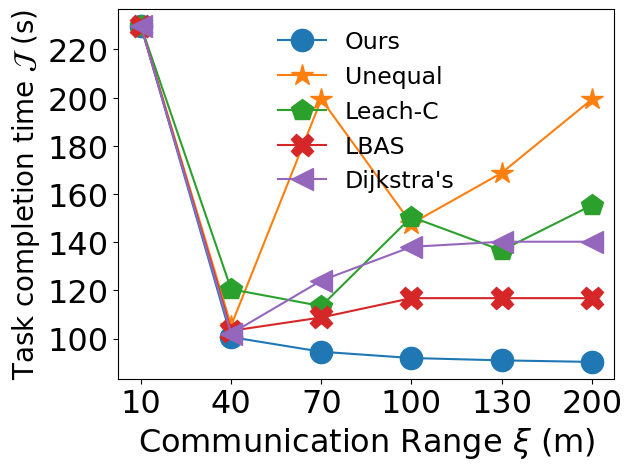}
    \label{fig:COMMRANGE}}
\vspace{-0.3cm}
\caption{(a) Illustration of the network with  $\xi=10$m. "$\boldsymbol{\times}$" marks the \textit{master} and the dashed circles indicates the communication range; (b) Performance comparison for different values of $\xi$.} \label{fig: ex1}
\vspace{-0.3cm}
\end{figure}

In the second experiment, we compare the performance of different approaches under varying server communication ranges. Specifically, we consider $N = 20$ servers distributed over the simulation area, as illustrated in Fig. \ref{fig:COMMRANGE_fig}. The communication range $\xi$ is configured to be the same for each server, and we vary $\xi$ from $10$m to $130$m. 
From Fig. \ref{fig:COMMRANGE}, we can see that our approach outperforms all benchmarks and continues to improve as the communication range $\xi$ increases. This improvement occurs because, as  $\xi$ grows, network connectivity strengthens as more servers within each other’s communication range. This expanded connectivity enables more servers to contribute to task processing, which, when properly selected, further reduces task completion time.

\section{Conclusion and Future Works} \label{sec:conclusion}
In this paper, we investigated the design of network topology to improve computational efficiency for task offloading in MEC. The proposed approach, DNTD-TO, draws inspiration from communication and sensor networks and builds three-layered network structures for task offloading in an iterative, decentralized manner. Additionally, it generates the optimal task allocation for the designed topology. To evaluate its performance, we conducted various comparison studies. The simulation results show that DNTD-TO significantly outperforms existing topology design methods. In the future, we will explore network structures with more than three layers and consider networks with mobile servers.


\section*{Acknowledgment}
We would like to thank the National Science Foundation under Grant CAREER-2048266 and CCRI-1730675 for the support of this work.

{\appendix[Proof of Lemma \ref{Thm:theorem 1}]
It is straightforward that when $C_i = \emptyset$, we have $J^* = \alpha_iy$. When $C_i \neq \emptyset$, we can solve problem $\mathcal{P}_1$ by relaxing it to a linear programming problem as follows:
\begin{align*}
    \mathcal{P}_2:&\min z \\
    s.t. \quad & z \geq y_l \alpha_l, l \in C_i \cup \{i\}  \\
    & \quad \quad y_0 + \sum_{l \in C_i} y_l = y
\end{align*}

To solve problem $\mathcal{P}_2$, the Lagrangian multiplier method \cite{10622383} can be used, with the Lagrangian function given by 
 $   \mathcal{L} = z + \sum_{l \in C_i \cup \{i\}}\lambda_l(y_l \alpha_l - z) + \mu(\sum_{l \in C_i \cup \{i\} }y_l - y)$, 
where $\boldsymbol{\lambda} = \{ \lambda_l \geq 0 | l \in C_i \cup \{i\} \}$ and $\mu$ are Lagrangian multipliers. 
As it fulfills the Slater's condition \cite{auslender2000lagrangian}, we can resort to the Karush-Kuhn-Tucker (KKT) condition \cite{luo2006introduction} to solve this problem:
\begin{subnumcases} {\label{eq: kkt}}
    \frac{\partial}{\partial y_l}\mathcal{L} = 0, ~ \forall l \in C_i \cup \{i\}  \label{eq: first}\\
    \frac{\partial}{\partial z}\mathcal{L}  = 0 \label{eq: second} \\
    \lambda_l (y_l \alpha_l - z) = 0, ~\forall l \in C_i \cup \{i\}  \label{eq: third} \\
    \sum_{l \in C_i \cup \{i\}} y_l = y \notag \\
    y_l \alpha_l - z \leq 0, ~\forall l \in C_i \cup \{i\}  \notag \\
    \lambda_l \geq 0, ~\forall l \in C_i \cup \{i\}  \notag
\end{subnumcases}

If $\alpha_ly^*_l \neq \alpha_iy^*_i$ and let
$j=\argmax_{l \in C_i \cup \{i\}} \alpha_l y^*_l$, then (\ref{eq: first}) - (\ref{eq: third}) can be rewritten as:
\begin{subnumcases} {}
    \frac{\partial}{\partial y_j} \mathcal{L} = \lambda_j \alpha_j + \mu = 0 \label{eq: 1} \\
    \frac{\partial}{\partial z} \mathcal{L} = 1 - \sum_{l \in C_i \cup \{i\} } \lambda_l = 0 \label{eq: 2} \\
    \lambda_l (\alpha_l y_l - z) = 0, \forall l \neq n, l \in C_i \cup \{i\}  \label{eq: 3} 
\end{subnumcases}
Since $z = \alpha_j y^*_j$, we can derive from (\ref{eq: 3}) that $\lambda_l = 0,\forall l \neq j, l \in C_i \cup \{i\}$, from (\ref{eq: 2}) that $\lambda_j = 1$, and then from (\ref{eq: 1}) that $\mu = -\alpha_j < 0$. 
However, for $\forall l \neq j$, (\ref{eq: first}) can also be written as 
$\frac{\partial}{\partial y_l} \mathcal{L} = \lambda_l \alpha_l + \mu = 0 = \mu$. 
This produces a conflict value for $\mu$. Therefore, the optimal task allocation has to satisfy $\alpha_ly^*_l = \alpha_iy^*_i, \forall l \in C_i$, and $J^* = \alpha_ly^*_l = \alpha_iy^*_i$.

}

\bibliographystyle{IEEEtran}
\bibliography{reference}

\end{document}